\documentclass[11pt]{article} \usepackage[dvips]{epsfig}

\def\ds{\displaystyle} \def\bb{\bibitem} \def\lb{\label}
\def\be{\begin{equation}} \def\ee{\end{equation}}
\def\ba{\begin{eqnarray}} \def\ea{\end{eqnarray}} \def\part{\partial}
\def\L{\Lambda}  \def\s{\sigma} \def\a{\frac1{c^2+1}}
\def\ka{\frac{k}{c^2+1}} \def\tr{\tilde{r}} \def\tp{\tilde{\phi}}
\def\ts{\tilde{\s}}

\begin{document}

\date{1 September 2001}

\begin{titlepage}
\title{Critical collapse in 2+1 dimensional AdS spacetime:\\
quasi-CSS solutions and linear perturbations} \author{G\'erard
Cl\'ement$^{a}$ \thanks{Email: gclement@lapp.in2p3.fr} and Alessandro
Fabbri$^{b}$ \thanks{Email: fabbria@bo.infn.it} \\ \\ {\small
$^{(a)}$Laboratoire de  Physique Th\'eorique LAPTH (CNRS),} \\ {\small
B.P.110, F-74941 Annecy-le-Vieux cedex, France}\\ {\small
$^{(b)}$Dipartimento di Fisica dell'Universit\`a di Bologna} \\
{\small and INFN sezione di Bologna,} \\ {\small Via Irnerio 46, 40126
Bologna, Italy} }

\maketitle

\begin{abstract}
We construct a one-parameter family of exact time-dependent solutions to 2+1 gravity with a negative cosmological
constant and a massless minimally coupled scalar field as source. These solutions present a continuously
self-similar (CSS) behavior near the central singularity, as observed in critical collapse, and an asymptotically
AdS behavior at spatial infinity. We consider the linear perturbation analysis in this background, and discuss the
crucial question of boundary conditions. These are tested in the special case where the scalar field decouples and
the linear perturbations describe exactly the small-mass static BTZ black hole. In the case of genuine scalar
perturbations, we find a growing mode with a behavior characteristic of supercritical collapse, the spacelike
singularity and apparent horizon appearing simultaneously and evolving towards the AdS boundary. Our boundary
conditions lead to the value of the critical exponent $\gamma = 0.4$.
\end{abstract}

\end{titlepage}
\setcounter{page}{2}
\section{Introduction}

The exploration of possible connections between seemingly different
fields of physics has always attracted much interest and curiosity. In
this respect, the observation first made by Choptuik \cite{chop} that
gravitational collapse does exhibit many of the features typical of
critical phenomena has opened a new of line of research (see for
instance the review paper \cite{gund} and references
therein). Specifically, Choptuik showed numerically, on the example of
a gravitating spherically symmetric massless scalar field, that
certain field configurations lying at the black hole threshold in the
phase space of initial data for the gravity-matter system are
characterized by universality, power-law scaling of the black hole
mass and self-similarity (discrete self-similarity in this specific
case). Recently, Pretorius and Choptuik \cite{CP} have considered the
same problem in a simplified context, namely the collapse of a
massless and minimally coupled scalar field in 2+1 dimensional AdS
spacetime.  The inclusion of a negative cosmological constant is
necessary for the existence of vacuum black hole solutions in 2+1
dimensions, the BTZ black holes \cite{BTZ}. By considering numerical
collapsing configurations close to black hole formation, they obtained
threshold solutions exhibiting characteristic critical features,
namely power-law scaling and continuous self-similarity
(CSS). Numerical calculations along similar lines were also performed
by Husain and Olivier \cite{HO}.

While the numerical investigation of critical collapse is by now a well-established branch of general relativity,
far less is known on the purely analytical level. In 3+1 dimensions, a family of exact CSS solutions to the
Einstein-massless scalar field model, first given by Roberts \cite{rob}, was studied by various
authors\cite{bra}-\cite{hay}. In particular, following the general analysis of linear perturbations of critical
solutions \cite{kha,maison}, Frolov studied the linear perturbations of the threshold Roberts solution, and
determined the mass-scaling exponent. In 2+1 dimensions, Garfinkle \cite{gar} found a class of CSS solutions to
the gravitating massless scalar field model with vanishing cosmological constant, and argued that one of these
solutions should approximate the critical solution observed in \cite{CP} near the singularity. The drawback is
that it is not possible to show that linear perturbations of such a solution lead to black hole formation, both
because the solution of the perturbation problem necessitates boundary conditions which cannot be consistently
stated for a solution which is known only in the vicinity of the singularity, and more fundamentally because black
hole formation in 2+1 dimensions requires a cosmological constant, which does not appear in Garfinkle's CSS
solutions. Subsequently, the present authors \cite{crit} found another CSS solution to the same problem, and
showed how to extend this solution to a quasi-CSS solution of the full problem with cosmological constant. It was
then possible to perform the linear perturbation analysis of this extended solution, and to show that it did lead
to black hole formation.

In this paper we generalize the work of \cite{crit} to a new one-parameter class of CSS solutions, and
significantly refine the analysis of black-hole formation. In section 2 we review Garfinkle's one-parameter family
of exact CSS solutions to the $\Lambda=0$ equations of motion. The question of extending these to solutions of the
full $\Lambda\neq 0$ system is addressed in section 3, where for a special parameter value we find an exact
cosmological solution. Then, in sections 4 and 5, we use a limiting procedure to derive from Garfinkle's solutions
a new  class of $\L = 0$ CSS solutions, discuss their properties, and carry out their extension to solutions for
the case with negative cosmological constant. The main contribution of this paper is represented by sections 6, 7
and 8, where we consider the linear perturbation analysis in the background of these quasi-CSS solutions. The
discussion of the most appropriate boundary conditions to be imposed on the perturbations, a separate analysis for
the limiting case where the scalar field decouples, and a discussion of two possible scenarios for black hole
formation in the linear approximation are then carried out. Finally, in section 9 we compare our results with
those of previous numerical and analytical investigations.

\setcounter{equation}{0}
\section{First class of CSS solutions}

The Einstein equations with a cosmological constant and a minimally
coupled massless scalar field source are \be\lb{ein} G_{\mu\nu} - \L
g_{\mu\nu} = \kappa T_{\mu\nu}\,, \ee where $T_{\mu\nu}$ is the
stress-energy tensor for the scalar field \be T_{\mu\nu} =
\part_{\mu}\phi\part_{\nu}\phi - \frac12 g_{\mu\nu}
\part^{\lambda}\phi\part_{\lambda}\phi\,, \ee the cosmological
constant $\L = -l^{-2}$ is negative for asymptoticaly AdS spacetime,
and the gravitational constant $\kappa$ is assumed to be positive.

The (2+1)-dimensional rotationally symmetric line element may be
written in double null coordinates as: \be\lb{an} ds^2 = e^{2\s}dudv -
r^2d\theta^2, \ee with metric functions $\s(u,v)$ and $r(u,v)$. The
Einstein equations (\ref{ein}) (where we take $\kappa = 1$) and the
associated scalar field equation are \ba & r_{,uv} = \frac{\ds\L}{\ds
2} r e^{2\s}, & \lb{ein1} \\ & 2\s_{,uv} = \frac{\ds \L}{\ds 2}
e^{2\s} - \phi_{,u}\phi_{,v}, & \lb{ein2} \\ & 2\s_{,u} r_{,u}
-r_{,uu} = r\phi_{,u}^2, & \lb{ein3} \\ & 2\s_{,v} r_{,v}  -r_{,vv} =
r\phi_{,v}^2, & \lb{ein4}\\ & 2r\phi_{,uv} + r_{,u}\phi_{,v} +
r_{,v}\phi_{,u} = 0 & \lb{phi}\,.  \ea For a given solution of these
equations, the Ricci scalar is \be\lb{risca} R = -6\L +
4e^{-2\s}\phi_{,u}\phi_{,v}.  \ee

Assuming $\L = 0$, Garfinkle has found \cite{gar} the following family
of exact CSS solutions to these equations \ba\lb{gar1} ds^2 & = &
-A\left(\frac{(\sqrt{\hat{v}} +
\sqrt{-\hat{u}})^4}{-\hat{u}\hat{v}}\right)^{c^2}\,d\hat{u}\,d\hat{v}
- \frac14(\hat{v}+\hat{u})^2\,d\theta^2\,, \nonumber \\ \phi & = &
-2c\ln(\sqrt{\hat{v}} + \sqrt{-\hat{u}})\,, \ea depending on an
arbitrary constant $c$ and a scale $A > 0$. These solutions are
continuously self-similar with homothetic vector
$(\hat{u}\part_{\hat{u}} + \hat{v}\part_{\hat{v}})$. An equivalent
form of these CSS solutions, obtained by making the transformation
\be\lb{q} -\hat{u} = (-\bar{u})^{2q}\,, \quad \hat{v} = (\bar{v})^{2q}
\qquad (1/2q = 1 - c^2) \ee to the barred null coordinates
$(\bar{u},\bar{v})$, is \ba\lb{gar2} ds^2 & = & -\bar{A}(\bar{v}^q +
(-\bar{u})^q)^{2(2q-1)/q}\,d\bar{u}\,d\bar{v} - \frac14(\bar{v}^{2q} -
(-\bar{u})^{2q})^2\,d\theta^2\,, \nonumber \\ \phi & = &
-2c\ln(\bar{v}^q + (-\bar{u})^q)\,.  \ea

Garfinkle suggested that the line element (\ref{gar1}) describes
critical collapse in the sector $r = -(\hat{u}+\hat{v})/2 \ge 0$, near
the future point singularity $\hat{u} = \hat{v} = 0$, where the effect
of the cosmological constant can be neglected. The global spacetime
structure of depends on the value of the parameter $c^2$ and is
discussed in \cite{crit}. As shown in \cite{gar}, the metric
(\ref{gar2}) can be extended through the surface $\bar{v} = 0$ only
for $q = n$, where $n$ is a positive integer.

Let us mention here that these solutions belong to a larger class of solutions which may be generated from the
Garfinkle solutions (\ref{gar1}) or (\ref{gar2}) by the Ida-Morisawa transformation. Ida and Morisawa considered
\cite{ida} three-dimensional Einstein-scalar fields with a hypersurface orthogonal spacelike Killing vector,
parametrized by (\ref{an}), and showed that the transformation
\be\lb{idamo} (r', \sigma', \phi') = (r, \sigma + b\phi + (b^2/2)\ln r, \phi + b\ln r) \ee generates a family
(parametrized by $b$) of Einstein-scalar fields. Applied to (\ref{gar2}), this transformation leads to the
solutions
\ba\lb{gar3} ds'^2 & = & -\bar{A}'(\bar{v}^q +
(-\bar{u})^q)^{(2c-b)^2}(- \bar{v}^q + (-\bar{u})^q)^{b^2}\,d\bar{u}\,d\bar{v} - \frac14(\bar{v}^{2q} -
(-\bar{u})^{2q})^2\,d\theta^2\,, \nonumber \\ \phi' & = & (b-2c)\ln(\bar{v}^q + (-\bar{u})^q) + b\ln(\bar{v}^q -
(-\bar{u})^q)\,, \ea with $q = 1/2(1-c^2)$. In the case $b=2c$, this solution (which was previously given in
\cite{crit}, Eq. (2.16)) is again CSS. For $b \neq   0$, these solutions have a timelike central ($r = 0$)
curvature singularity at $\bar{v}^q - (-\bar{u})^q = 0$.  They can again be extended through the surface $\bar{v}
= 0$ for $q = n$. In the case $q = n$ odd, the extension of a solution (\ref{gar3}) with parameters $(\bar{A}, b,
c)$ leads to a solution of the same class with parameters $(\bar{A}' = -\bar{A}, b' = b-2c, c' = -c)$.

\setcounter{equation}{0}
\section{A class of quasi-CSS cosmologies}

For a consistent analytical treatment of critical collapse, we should
be able to extend the $\L = 0$ CSS solutions (\ref{gar1}) to $\L < 0$
quasi-CSS solutions, i.e. to construct (explicitly or implicitly)
solutions of the full field equations with the behavior (\ref{gar1})
near the singularity. However, we have been unable to find an ansatz
which separates the variables and reduces the field equations to
ordinary differential equations, except in the special case $c^2 =
1/2$ ($q = 1$). In this case the solution (\ref{gar2}) with $\bar{A} =
1$ may be written, after transforming to coordinates ($R,T$) defined
by \be \bar{u} = \frac{R-T}{\sqrt2}, \quad \bar{v} =
\frac{R+T}{\sqrt2}, \ee in the form of a FRW cosmology with flat
spatial sections, \ba\lb{q1} ds^2 & = & T^2(\,dT^2 - dR^2 -
R^2\,d\theta^2), \nonumber \\ \phi & = & - \sqrt2\ln T.  \ea

This form suggests searching for an extension by separating the
equations in the variables $R, T$. With the metric parametrization
\be\lb{an1} ds^2 = e^{2\Sigma}(dT^2-dR^2) - r^2d\theta^2, \ee these
equations read \ba\lb{TR} \ddot{r} -r'' & = & 2\L r e^{2\Sigma},
\nonumber \\ 2(\ddot{\Sigma} - \Sigma'') & = & 2\L e^{2\Sigma} -
\dot{\phi}^2 + \phi'^2, \nonumber \\ 2(\dot{\Sigma}\dot{r}+\Sigma'r')
-\ddot{r} -r'' & = & r(\dot{\phi}^2 + \phi'^2) , \\ \Sigma'\dot{r} +
\dot{\Sigma}r' -\dot{r}' & = & r\dot{\phi}\phi', \nonumber\\
r(\ddot{\phi}-\phi'') + \dot{r}\dot{\phi} - r'\phi' & = & 0, \nonumber
\ea with $\dot{} = \partial/\partial_T$ and $' = \partial/\partial_R$.
Let us make the ansatz \be \phi = A(T) + B(R), \quad r = F(T)G(R).
\ee The equations separate and lead to a FRW cosmology if \be G'' = k
G \ee ($k$ constant). One finds that for the solution to reduce to
(\ref{q1}) for $\L =0$ one must have $B$ constant, e.g. $B = 0$, the
remaining equations being integrated by \ba\lb{cosm1} & & \dot{A} =
\frac{\alpha}{F}, \quad e^{\Sigma} = \beta F, \nonumber \\ & &
\dot{F}^2 - \alpha^2/2 - kF^2 - \L\beta^2 F^4 = 0, \ea where $\alpha$
and $\beta$ are integration constants; without loss of generality we
may choose $\alpha = \sqrt{2}$ (by rescaling times) and $\beta =
1$. The resulting metric \be\lb{cosm2} ds^2 = F^2(T)(dT^2 - dR^2
-G^2(R)d\theta^2), \ee where $G(R)$ solves the equation $G'' = kG$
with the initial conditions $G(0) = 0$, $G'(0) = 1$ for regularity, is
of the FRW type, and reduces to (\ref{q1}) in the limit $\L \to 0$ if
$k \to 0$ in this limit.

The discriminant $(k^2 - 4\L)$ of the effective potential in the last
equation (\ref{cosm1}) is positive definite for $\L$ negative, so that
this potential has two roots $F_-^2 = \tau_- < 0$ and $F_+^2 = \tau_+
> 0$. It follows that this cosmology is time symmetric around the
turning point $F^2 = \tau_+$, with initial and final singularities at $F^2 = 0$. Defining the new time variable
$\tau \equiv F^2(T)$, the line element (\ref{cosm2}) may be rewritten as\footnote{For $G(R) =$ const. ($k = 0$, so
that $\tau_- = -\tau_+$), this solution reduces to the cosmological version of the static solution (\ref{cicca})
with $a = 0$ (put $\rho_- = -\rho_+ = -\tau_+$ and $\rho^2 = \tau_+^2 - \tau^2$).}  \be\lb{cosm3} ds^2 =
\frac{l^2}4 \frac{d\tau^2}{(\tau-\tau_-)(\tau_+-\tau)} - \tau (dR^2 + G^2(R)d\theta^2), \ee ($\L = -l^{-2}$) with
\ba G(R) & = & \mu^{-1}\sinh\mu R \quad (k = \mu^2 > 0), \nonumber \\ G(R) & = & R \quad (k = 0), \\ G(R) & = &
\nu^{-1}\sin\nu R \quad (k = -\nu^2 < 0). \nonumber \ea In the special case $k = 0$, this line element may be
rewritten as \be ds^2 = F^2\left(\frac{dF^2}{1-F^4/l^2} - dR^2 - R^2d\theta^2\right), \ee showing clearly the
restoration of self-similarity in the limit $l \to \infty$.

\setcounter{equation}{0}
\section{Second class of CSS solutions}

From the Garfinkle class of CSS solutions (\ref{gar1}), we may derive
a new class of CSS solutions by a generalization of the construction
of \cite{crit}. The double null ansatz (\ref{an}) is invariant under
boosts $\hat{u} \to \alpha \hat{u}$, $\hat{v} \to
\alpha^{-1}\hat{v}$. Rescaling angles by $\theta \to (2/\alpha)\theta$
and putting $A = \alpha^{-2c^2}\hat{A}$, we obtain from (\ref{gar1}),
in the limit of an infinite boost $\alpha \to \infty$, \ba\lb{new1}
ds^2 & = & -\hat{A}(\frac{-\hat{u}}{\hat{v}})^{c^2}d\hat{u}d\hat{v} -
\hat{u}^2 d\theta^2, \nonumber \\ \phi & = & -c \ln(-\hat{u}).  \ea
These new solutions are obviously, as the original Garfinkle
solutions, continuously self-similar with homothetic vector
$(\hat{u}\part_{\hat{u}} + \hat{v}\part_{\hat{v}})$.  A simpler form
of these CSS solutions may be obtained by transforming to new null
coordinates \be u = - (-\hat{u})^{1+c^2}, \quad v = \hat{v}^{1-c^2},
\ee which yields (with the normalization $\hat{A} = c^4-1$)
\ba\lb{new2} ds^2 & = & dudv - (-u)^{2/(1+c^2)} d\theta^2, \nonumber
\\ \phi & = & -\frac{c}{1+c^2}\ln(-u).  \ea For $c^2 = 1$ we recover
the ``new CSS solution'' of \cite{crit}.

Interestingly, the CSS solutions (\ref{new1}) may also be obtained by
submitting the singular static solutions of the cosmological
Einstein-scalar field equations to an infinite boost near the
singularity. These solutions, derived in \cite{sigcosm}, \ba ds^2 & =
& A\,|\rho - \rho_+|^{1/2+a}\,|\rho - \rho_-|^{1/2-a}\,dt^2 -
\,\frac{4|\L|}{A}\,|\rho - \rho_+|^{1/2-a}\,|\rho - \rho_-|^{1/2+a}\,
d\theta^2 \nonumber \\ & & + \,\frac{d\rho^2}{4\L(\rho - \rho_+)(\rho
- \rho-)}\,, \qquad \phi = \sqrt{\frac{1-4a^2}8}\,
\ln\left(\frac{|\rho-\rho_+|}{|\rho-\rho_-|}\right)
\label{cicca}
\ea (with $-1/2 \le a \le 1/2$, and $\rho_{\pm} = O(\L^{-1}$)) are
singular for $\rho = \rho_{\pm}$.  In the neighborhood of one of these
singularities, e.g. $\rho_+$, the transformation $\rho = \rho_+ +
x^{4/(1-2a)}$ leads (after appropriate rescalings) for $x \to 0$, or
equivalently for $\L \to 0$, to the exact $\L = 0$ solution \cite{sig}
\be ds^2 = x^{2(1+2a)/(1-2a)}(dt^2-dx^2) - x^2 d\theta^2, \quad \phi =
\sqrt{\frac{2(1+2a)}{1-2a}}\ln x.  \ee Transforming to null
coordinates $\hat{u} = t -x$, $\hat{v} = t + x$, and boosting this
solution by $\hat{u} \to \alpha \hat{u}$, $\hat{v} \to
\alpha^{-1}\hat{v}$, we obtain in the limit of an infinite boost
(again after appropriate rescalings) \be ds^2 =
(-\hat{u})^{2(1+2a)/(1-2a)}d\hat{u}d\hat{v} - \hat{u}^2 d\theta^2,
\quad \phi = \sqrt{\frac{2(1+2a)}{1-2a}}\ln|\hat{u}|.  \ee After a
suitable transformation on the coordinate $\hat{v}$, this is found to
coincide with the CSS solution (\ref{new1}) provided we identify \be
c^2 = 2\frac{1+2a}{1-2a}.  \ee

For $a = -1/2$ ($+1/2$) the solution (\ref{cicca}) is no longer
singular for $\rho = \rho_+$ ($\rho_-$) but reduces to the BTZ black
hole \be\lb{btz} ds^2 = (r^2/l^2-M)dt^2 - \frac{dr^2}{(r^2/l^2-M)} -
r^2d\theta^2, \ee with $\phi = 0$. Near $r = 0$, this reduces for $M >
0$ to the Minkowski cosmology $ds^2 = M^{-1}dr^2 - Mdt^2
-r^2d\theta^2$,  which after an infinite boost leads to the metric
(\ref{new2}) with $c = 0$, \be\lb{new0} ds^2 = dudv - u^2 d\theta^2.
\ee The infinite boost limit corresponds physically to the massless
limit, and so one would expect that the metric (\ref{new0}) can
alternatively be derived from (\ref{btz}) by taking the limit $M \to
0$ (this time without an infinite boost). Indeed, the extreme ($M =
0$) BTZ black hole, or ``vacuum solution'' \be\lb{vac} ds^2 =
(r^2/l^2)dt^2 - \frac{dr^2}{r^2/l^2} - r^2d\theta^2, \ee may be
rewritten (with $x = l^2/r$) as \ba\lb{vac1} ds^2 & = &
\frac{l^2}{x^2}(dt^2- dx^2 - l^2d\theta^2) \nonumber \\ & = &
\frac{4l^2}{(U-V)^2}(dUdV - l^2d\theta^2) \ea in null coordinates $U =
t-x$, $V= t+x$. The transformation \be U = -2l^2/u,\quad V = v/2 \ee
transforms (\ref{vac1}) to \be\lb{vac2} ds^2 =
\frac1{(1+uv/4l^2)^2}(dudv - u^2d\theta^2).  \ee This goes over to the
$c^2 = 0$ CSS solution (\ref{new0}) in the limit $\L \to 0$ ($l \to
\infty$) or, equivalently, in the near-horizon limit $r \to 0$ ($u \to
0$).

The $c^2 = 0$ spacetime (\ref{new0}) is Ricci-flat, and so (in 2+1
dimensions) is flat. Therefore it is locally equivalent to Minkowski
spacetime. The explicit local transformation from the Minkowski metric
written in the double null form \be ds^2 = du\,dw - dy^2 \ee to the
metric (\ref{new0}) is \be u = u, \quad y = u\theta, \quad w = v +
u\theta^2.  \ee The rotationally symmetric $c^2 = 0$ CSS spacetime is
obtained by periodically identifying $\theta$ with $\theta +
2\pi$. From this vacuum solution ($r = -u$, $\sigma= 0$, $\phi = 0$),
the entire second class (4.1) of $c^2 \neq 0$ CSS solutions may be
generated by the Ida-Morisawa transformation (\ref{idamo}) with $b =
c$.

For all $c \neq 0$, the CSS solutions (\ref{new2}) have a high degree
of symmetry, with the four Killing vectors \ba\lb{kill} L_1 & = &
v\part_{v} - u\part_{u} + \frac1{c^2+1}\,\theta\part_{\theta}\,,
\nonumber \\ L_2 & = & -2\frac{c^2-1}{c^2+1}\,\theta\part_{v} +
(-u)^{\frac{c^2-1}{c^2+1}}\,\part_{\theta}\,, \nonumber \\ L_3 & = &
\part_{v}\,, \nonumber \\ L_4 & = & \part_{\theta}\,.  \ea For $c^2 =
1$, the above form of $L_2$ is replaced by \be L_2 = \theta\part_{v}
-\frac12\ln(-u)\part_{\theta}.  \ee As in the case $c^2 =1$
\cite{crit}, the Lie algebra generated by the Killing vectors
(\ref{kill}) is solvable. In the case $c^2 = 0$, the flat metric
(\ref{new0}) admits of course six Killing vectors generating the
(2+1)-dimensional Poincar\'e algebra.

Again as in the case $c^2 = 1$, the metric (\ref{new2}) is devoid of
scalar curvature singularities. The nature of the corresponding
geometry may be understood from the study of geodesic motion in this
spacetime. The geodesic equations are integrated by  \be\lb{geos}
\dot{u} = \pi\,, \quad (-u)^{2/(c^2+1)}\dot{\theta} = -l\,, \quad
\pi\dot{v} + l\dot{\theta} = \varepsilon\,,  \ee  where $\pi$, $l$ and
$\varepsilon$ are constants of the motion. The spacetime is extendible
to a geodesically complete spacetime in two cases:

a) $c^2 = 0$. This is the near-horizon limit (\ref{new0}) of the BTZ
vacuum (\ref{vac2}), and is diffeomorphic to the Minkowski metric, as
explained above.

b) $c^2 \to \infty$. In this limit the scalar field decouples and the
metric (\ref{new2}) reduces to the geodesically complete cylindrical
Minkowski metric.

In the other cases, the third equation (\ref{geos}) integrates to
\be\lb{geos1} v = \frac{\varepsilon}{\pi^2}u -
\frac{l^2(c^2+1)}{\pi^2(c^2-1)}(-u)^{(c^2-1)/(c^2+1)} + {\mathrm
const.}\,,  \ee  showing that for $c^2 < 1$ nonradial geodesics
terminate at $u = 0,  v \to +\infty$, while radial geodesics ($l = 0$)
can be continued through the null line $u = 0$, which they cross at
finite  $v$, to $u \to +\infty$, only the endpoint $v \to +\infty$ of
the null line $u = 0$ being singular.  In the case $c^2 = 1$ (treated
in \cite{crit}), (\ref{geos}) is replaced by \be  v =
\frac{\varepsilon}{\pi^2}u - \frac{l^2}{\pi^2}\ln(-u) + {\mathrm
const.}\,, \ee  leading to the same conclusion as for $c^2 < 1$. On
the other hand,  for $c^2 > 1$  nonradial geodesics terminate at $u =
0$, $v =$ const.,  and the whole line $u = 0$ corresponds to a
singularity of the geometry. So the $c^2 = 1$ solution is an extreme
solution in the manifold of CSS  solutions (\ref{new2}), lying at the
threshold between two classes of  solutions ($0 < c^2 \le 1$ and $c^2
> 1$) differing by their global spacetime geometry (Fig. 1).

\setcounter{equation}{0}
\section{Quasi-CSS extensions of the second class}

In this section we proceed to extend the second class of $\L = 0$ CSS
solutions (\ref{new2}) to exact solutions of the full $\L < 0$
equations (the self-similarity being then broken by the cosmological
constant). Following \cite{crit} we make the ansatz \be\lb{anext} ds^2
= e^{2\nu(x)}dudv - (-u)^{\frac2{c^2+1}}\rho^2(x)d\theta^2, \quad \phi
= -\frac{c}{c^2+1}\ln|u| + \psi(x)\,, \ee with $x = uv$. This reduces
to the CSS solution (\ref{new2}) for $\rho = 1, \, \nu = \psi = 0$,
and preserves the Killing subalgebra $(L_1, L_4)$. Inserting this
ansatz into the field equations (\ref{ein1})-(\ref{phi}) leads to the
system \ba x\rho'' + \frac{c^2+2}{c^2+1}\rho' & = & \frac{\L}{2}\rho
e^{2\nu}, \lb{einx1}\\ 2(x\nu'' + \nu') +\psi'(x\psi'-\frac{c}{c^2+1})
& = & \frac{\L}{2}e^{2\nu}, \lb{einx2}\\ x^2(-\rho'' + 2\rho'\nu'-
\rho\psi'^2) + \frac2{c^2+1}x(- \rho' + \rho(\nu'+c\psi')) & = & 0
\lb{einx3} \\ - \rho'' + 2\rho'\nu'- \rho\psi'^2 & = & 0 \lb{einx4} \\
2x(\rho\psi')' + \frac{2c^2+3}{c^2+1}\rho\psi' & = &
\frac{c}{c^2+1}\rho' \lb{phix} \ea ($' = d/dx$). A simple first
integral, obtained by comparing (\ref{einx3}) and (\ref{einx4}), is
\be\lb{firstint} \lb{rhospsi} \rho = e^{\nu+c\psi} \ee (with the
integration constant set to 1 by the boundary conditions
(\ref{boundx})).

The boundary conditions \be\lb{boundx} \rho(0) = 1, \quad \nu(0) = 0,
\quad \psi(0) = 0 \ee lead to a unique solution reducing to
(\ref{new2}) near $u = 0$. The small $x$ behavior of this solution is
\ba\lb{smallx} \rho & \simeq & 1 + \frac{c^2+1}{2(c^2+2)}\L x,
\nonumber\\ \nu & \simeq & \frac{(c^2+1)(c^2+3)}{2(c^2+2)(2c^2+3)}\L
x, \\ \psi & \simeq & \frac{c(c^2+1)}{2(c^2+2)(2c^2+3)}\L x. \nonumber
\ea

The analysis of \cite{crit} can be straightforwardly extended here,
leading to the conclusion that when $x$ decreases from $x = 0$, the
functions $\rho$ and $e^{2\nu}$ increase indefinitely, going to
infinity for a finite value $x = x_1$. The numerical solution of the
system (\ref{einx1})-(\ref{phix}) with boundary conditions
(\ref{boundx}) leads to values $x_1(c^2)$ equal to $-4l^2$ for $c^2 =
0$ and for $c^2 \to \infty$, and remaining of the order of $-4l^2$ for
finite $c^2$, the maximum value being $x_1(1) = -3.876 l^2$.  The
behavior of the metric functions near $x_1$ is found to be
\ba\lb{bound0} \rho & = & \rho_1\left(\frac1{\bar{x}} +
\frac{c^2}{2(c^2+1)x_1} -
\frac{c^2\bar{x}\ln(\bar{x})}{12(c^2+1)^2x_1^2} + ... \right)\nonumber
\\ e^{2\nu} & = & \frac{4x_1}{\L \bar{x}^2}\left(1 +
\frac{c^2\bar{x}^2\ln(\bar{x})} {12(c^2+1)^2x_1^2} + ...\right) \\
\psi & = & \psi_1 + \frac{c\bar{x}}{2(c^2+1)x_1} -
\frac{c\bar{x}^2}{8(c^2+1)^2x_1^2}\ln(\bar{x}) + ... \nonumber \ea
($\bar{x} = x - x_1$). Changing to coordinates ($\bar{T},\bar{R}$)
defined by \be u = -l^{c^2+1}e^{\bar{R}-\bar{T}}, \quad v =
l^{1-c^2}e^{\bar{R}+\bar{T}}, \ee we obtain from (\ref{bound0}) the
leading asymptotic behaviors near $\bar{R} = \bar{R}_1$ \be ds^2
\simeq \frac{l^2}{(\bar{R}_1-\bar{R})^{2}}(d\bar{T}^2-d\bar{R}^2 -
e^{\frac2{c^2+1}(\bar{T}_1 -\bar{T})}d\theta^2)\,, \quad \phi = \phi_1
+ c\bar{T}/(c^2+1) \ee ($\bar{R}-\bar{R}_1\simeq
\bar{x}/2x_1$). Making the further coordinate transformation, \be
\bar{R}-\bar{R}_1 = -(c^2+1)/XT\,, \quad \bar{T}-\bar{T}_1 =
(c^2+1)\ln(T/(c^2+1))\,, \ee we arrive at the asymptotic form \be ds^2
\simeq l^2\left(X^2 dT^2 - \frac{dX^2}{X^2}  - X^2d\theta^2 \right),
\quad \phi = \phi_1 + c\ln T\,, \ee showing that the metric is
asymptotically AdS (with logarithmic subdominant terms). The
corresponding conformal diagrams for the cases $0 < c^2 \le 1$ and
$c^2 > 1$ are shown in Fig. 2.

For $c^2 = 0$ or $c^2 \to \infty$, the solution of Eq. (\ref{phix}) is
$\rho\psi' = \alpha x^{-p}$, with $\alpha$ constant and $p = 3/2$ for
$c^2 = 0$ or $p = 1$ for $c^2 \to \infty$, so that the boundary
conditions (\ref{boundx}) enforce $\alpha = 0$, i.e. $\psi =0$, and
the extended metric (\ref{anext}) is a vacuum metric. For $c^2 = 0$
this metric is \be ds^2 = \frac1{(1+uv/4l^2)^2}(dudv - u^2d\theta^2),
\ee which we recognize as the BTZ vacuum (\ref{vac2}). For $c^2 \to
\infty$, the extended metric \footnote{In this case the derivation of
(\ref{firstint}) breaks down. The solution (\ref{btz3}) is obtained by
first integrating (\ref{einx4}) to $\rho' =$ const. $e^{2\nu}$, then
integrating (\ref{einx1}) with the boundary conditions
(\ref{boundx}).}  \be\lb{btz3} ds^2 = \frac1{(1+uv/4l^2)^2}(dudv -
(1-uv/4l^2)^2d\theta^2), \ee is obtained from the BTZ metric
(\ref{btz}) by transforming first to coordinates $(R,T)$ with
\be\lb{coordout} r = M^{1/2}l\coth({M^{1/2}R/l}), \quad t = T, \ee
which leads to \be\lb{btz2} ds^2 =
\frac{M}{\sinh^2({M^{1/2}R/l})}(dT^2-dR^2) -
Ml^2\coth^2({M^{1/2}R/l})\,d\theta^2, \ee then, choosing the
convenient normalization $Ml^2 = 1$, to coordinates ($u,v$) defined by
\be u = -2l\,e^{(R-T)/l^2}, \quad v = 2l\,e^{(R+T)/l^2}.  \ee So the
family of time-dependent quasi-CSS solutions (\ref{anext})
interpolates between the BTZ vacuum for $c^2 = 0$ and the massive BTZ
metric for $c^2 \to \infty$.

\setcounter{equation}{0}
\section{Perturbations: general setup}

Now we study linear perturbations of the quasi-CSS solutions
(\ref{anext}) along the lines of the analysis of
\cite{fro,hay,crit}. The relevant parameter in critical collapse being
the retarded time \cite{gar} $T = -\ln(-\hat{u}) = -
(1/(c^2+1))\ln(-u)$, we expand these perturbations in modes
proportional to $e^{kT} = (-u)^{-k/(c^2+1)}$, with $k$ a complex
constant. Only those modes with $Re(k) > 0$, which grow when $T \to
+\infty$, will possibly lead to black hole formation. Keeping only one
mode, we decompose the perturbed fields as \ba\lb{pertan} r & = &
(-u)^{1/(c^2+1)}(\rho(x) + (-u)^{-k/(c^2+1)}\tr(x)), \nonumber\\ \phi
& = & -\frac{c}{c^2+1}\ln|u| + \psi(x) + (-u)^{-k/(c^2+1)}\tp(x), \\
\s & = & \nu(x) + (-u)^{-k/(c^2+1)}\ts(x). \nonumber \ea The
linearization of the Einstein equations (\ref{ein1})-(\ref{phi}) in
the perturbations $\tr$, $\tp$, $\ts$ leads to the system \ba & &
x\tr'' + \frac{c^2+2-k}{c^2+1}\,\tr' = \frac{\L}{2}\, e^{2\nu}(\tr +
2\rho\ts), \lb{pein1}\\ & & 2x\ts'' +\frac{2(c^2+1-k)}{c^2+1}\,\ts' =
\L e^{2\nu}\ts - \left(2x\psi'-\frac{c}{c^2+1}\right)\tp' +
\frac{k}{c^2+1}\,\psi'\tp, \lb{pein2}\\ & & -(1-k)x\tr' +
\left((1-k)x\nu'+\frac{k(1-k-c^2)}{2(c^2+1)}\right) \tr + \rho x\ts' -
k\left(x\rho' + \frac{\rho}{c^2+1}\right) \ts =  \nonumber \\ & &
\qquad \qquad \qquad -\rho\left(cx\tp' -
k\left(\frac{c}{c^2+1}-x\psi'\right)\tp\right) - cx\psi'\tr,
\lb{pein3}\\ & & 2(\rho'\ts' + \nu'\tr') - \tr'' = \psi'(2\rho\tp' +
\psi'\tr), \lb{pein4}\\ & & 2x\rho\tp'' + \left(2x\rho' +
\frac{2c^2+3-2k}{c^2+1}\, \rho\right)\tp' - \frac{k}{c^2+1}\,\rho'\tp
+ \left(2x\psi'- \frac{c}{c^2+1}\right)\tr' \nonumber \\ & & \qquad
\qquad \qquad + \left(2x\psi'' +
\frac{2c^2+3-k}{c^2+1}\,\psi'\right)\tr = 0.  \lb{pphi} \ea

This linear differential system is of order four (Eqs. (\ref{pein3})
and (\ref{pein4}) are constraints, while the perturbed scalar field
equation (\ref{pphi}) is a consequence of the perturbed Einstein
equations). An exact solution of the system (\ref{pein1})-(\ref{pphi})
is given by \ba\lb{gauge} \tr_k(x) & = & \alpha
(-x)^{1+k/(c^2+1)}\rho'(x)\,, \nonumber \\ \tp_k(x) & = & \alpha
(-x)^{1+k/(c^2+1)}\psi'(x)\,, \\ \ts_k(x) & = & \alpha
\left((-x)^{1+k/(c^2+1)}
\nu'(x)-\frac{c^2+1+k}{2(c^2+1)}(-x)^{k/(c^2+1)}\right)\,. \nonumber
\ea The corresponding first-order perturbed fields (\ref{pertan}) are
generated from the unperturbed fields by the gauge transformation $v
\to v - \alpha v^{1+k/(c^2+1)}$. So, up to gauge transformations, the
general solution of this system depends only on three integration
constants, which should be determined by enforcing appropriate
boundary conditions.

At this point we should emphasize that the question of which boundary
conditions are truly ``appropriate'' is rather subtle in the
general-relativistic context (for instance, in the related problem of
scalar field collapse in 3+1 dimensions, different boundary conditions
have been chosen in \cite{fro} and \cite{hay}), yet crucial for a
proper determination of the eigenmodes.

The three boundaries of our quasi-CSS solutions are $u =0$, $v = 0$
(or $x = 0$), and $x = x_1$ (the AdS boundary). At the boundary $u =
0$, corresponding to the original centre of the unperturbed quasi-CSS
solution, it seems natural to impose that (A) the perturbed metric
component $r$ does not diverge too quickly, which would conflict with
the linear approximation \cite{hay}. However this is subject to some
ambiguity. For instance, should one consider the limit $u \to 0$ with
$v$ held fixed, or rather with $x = uv$ held fixed? Also, the
perturbed spacetime may develop an apparent horizon hiding the
singularity at $u = 0$. In such a case it may seem preferable to view
this apparent horizon, rather than the line $u =0$, as the natural
boundary of the perturbed spacetime on which to impose a boundary
condition.

On the second boundary $v = 0$, it seems natural to require that (B1)
the perturbed solution matches smoothly the original quasi-CSS
solution \cite{fro} \be\lb{bound2} \tr(0) = 0, \quad \ts(0) = 0, \quad
\tp(0) = 0 \ee (on account of (\ref{boundx}), the perturbed solution
then also matches smoothly the original CSS solution).  However
conditions (\ref{bound2}) do not ensure that the curvature tensor
remains finite on $v = 0$, which seems essential for a smooth
matching. This last requirement is automatically satisfied if the
``weak'' conditions (\ref{bound2}) are supplemented with the the
``strong'' condition (B2) that the perturbations be (as the
unperturbed quasi-CSS solution) analytic in $v$ \cite{crit}.

Finally, we shall require that (C) the perturbations grow slowly
(i.e. not faster than the original fields) as the third boundary $x =
x_1$ is approached \cite{fro}, this condition ensuring the validity of
the linear approximation near the AdS boundary.  Note that, owing to
the gauge freedom, it is only necessary that these various boundary
conditions be satisfied up to gauge transformations, i.e. up to the
addition of a gauge perturbation (\ref{gauge}).

To enforce boundary conditions on the boundary $v = 0$, we must first
investigate the behavior of the solutions of the linearized system
near the boundary $x = 0$. Generalizing the analysis of \cite{crit},
we assume the power-law behavior \be\lb{pow} \tr(x) \propto (-x)^{p}
\ee ($p$ constant). For the purpose of the determination of the
exponent $p$, Eqs. (\ref{pein1}), (\ref{pein2}) and (\ref{pein4}) can
be approximated near $x = 0$ as \ba & & x\tr'' +
\frac{c^2+2-k}{c^2+1}\,\tr' \simeq \L\ts, \lb{pein01}\\ & & x\ts'' +
\frac{c^2+1-k}{c^2+1}\,\ts' \simeq \frac{c}{2(c^2+1)}\,\tp'
\lb{pein02}\\ & & \frac{c^2+1}{c^2+2}\,\L\ts' - \tr'' \simeq
\frac{c(c^2+1)}{(c^2+2)(2c^2+3)}\,\L\tp' \lb{pein04} \ea (in
(\ref{pein04}) we have replaced the unperturbed fields by their small
$x$ behaviors (\ref{smallx})). The discussion of this system is
simpler in the special (vacuum) case $c^2 = 0$, which we shall
consider in the next section, leaving the discussion of the general
(scalar) case $c^2 \neq 0$ to Sect. 8.

\setcounter{equation}{0}
\section{BTZ black holes as vacuum perturbations}

In the case $c^2 = 0$, we have seen that the quasi-CSS solution is the
BTZ vacuum, with $\psi = 0$. Then, the linearized equations
(\ref{pein1})-(\ref{pphi}) decouple into the linearized sourceless
Einstein equations and the linearized scalar field equation on the BTZ
vacuum background. The solution of the three-dimensional cosmological
Einstein equations being essentially unique, up to diffeomorphisms,
the three-parameter family of perturbations of the BTZ vacuum must
include the small-mass BTZ black holes, with the mass as perturbation
parameter (amplitude), as we now check explicitly.

When $c^2 = 0$, Eqs. (\ref{pein02}) and (\ref{pein04}) decouple and
must be satisfied separately. Then $\ts$ can be eliminated between
(\ref{pein01}) and (\ref{pein04}), leading to the third order equation
\be x\tr''' + (1-k)\tr'' \simeq 0, \ee which implies the secular
equation \be\lb{sec0} p(p-1)(p-1-k) = 0.  \ee We note that the root p
= 1+k corresponds to the gauge mode (\ref{gauge}). Discarding this
mode, we obtain from the full equations (\ref{pein1})-(\ref{pphi}) the
behavior of the perturbations near $x = 0$ in terms of two integration
constants $A$, $B$: \ba & &\tr(x) \sim A +B(-x), \\ & &\ts(x) \sim -
\frac{A}2 - (2-k)\L^{-1}B +O(x).  \ea This can only satisfy the
boundary conditions (\ref{bound2}) at $x = 0$  if \be\lb{btzb} A = 0,
\quad k = 2.  \ee It follows that in this case the solution of the
linearized system (\ref{pein1})-(\ref{pein4}) with the initial
conditions (\ref{bound2}) is, up to a gauge transformation, unique.
We now show that this solution is simply a linearization of the BTZ
metric.

The BTZ metric (\ref{btz2}) may be written in double-null form as
\be\lb{btz4} ds^2 = \frac{4l^2\,dU\,dV}{\sinh^2(U - V)} -
Ml^2\coth^2(U - V)\,d\theta^2, \ee with $U = \mu(T + R)$, $V = \mu(T -
R)$, and $\mu = M^{1/2}/2l$. This may again be rewritten, in terms of
the new null coordinates \be u = - 2\mu l^2\coth U, \quad v =
(2/\mu)\tanh V, \ee as \ba\lb{btz5} ds^2 & = &
\frac{\sinh^2U\cosh^2V}{\sinh^2(U-V)}\,du\,dv - 4\mu^2l^4
\coth^2(U-V)\,d\theta^2 \nonumber \\ & = & \frac1{(1+uv/4l^2)^2}(dudv
- (u+(M/4)v)^2d\theta^2).  \ea This form of the BTZ metric again
reduces to the vacuum CSS metric (\ref{new0}) in the infinite boost
limit (which amounts to taking $v \to 0$) and, furthermore, can
obviously be considered for small masses as a perturbation of the BTZ
vacuum (\ref{vac2}). From (\ref{btz5}) the metric function $r$ is
\ba\lb{pvac} r & = & \rho\,(-u-(M/4)v) \nonumber\\ & = &  - u\,(\rho +
(M/4)u^{-2}x\rho) , \ea with $x = uv$, and $\rho(x) =
(1+x/4l^2)^{-1}$. This is of the form of (\ref{pertan}) for the
eigenmode $k = 2$, as found above.  On the other hand, the metric
function $e^{2\s}$ in (\ref{btz5}) does not depend on the perturbation
parameter $M/4$, so that the perturbations $\tr$ and $\ts$ are simply
given by \be\lb{btzp} \tr = \frac{M}4 x\rho, \quad \ts = 0, \ee which
satisfy the condition of slow growth at the AdS boundary $x \to -4l^2$
(where $\tr \sim -Ml^2\rho$), as well as the requirement of finiteness
of $r$ for $u \to 0$ with $v$ held fixed (but not for $u \to 0$ with
$x$ held fixed), and are easily checked to solve exactly the vacuum
system (\ref{pein1})-(\ref{pein4}).

It is instructive to analyze more closely in this (obviously very
special) case the phenomenon of black hole formation. First we note
that the unperturbed metric (\ref{vac2}) can be extended beyond $u =
-\infty$ by the coordinate transformation $u = -\bar{u}^{-1}$, leading
to the metric \be\lb{new0i} ds^2 =
\frac{1}{(\bar{u}-v/4l^2)^2}(d\bar{u}dv - d\theta^{2}), \ee which is
regular in the triangle $\bar{u}-v/4l^2 > 0$, the other boundaries of
this triangle being the apparent horizons $r_{,v}=0$ ($u = 0$) and
$r_{,u}=0$ ($v = -\infty$) through which the metric can be
periodically extended to the geodesically complete BTZ vacuum
spacetime (Fig. 3). Bearing in mind that the perturbed metric
(\ref{pvac}) can similarly be extended beyond $u = -\infty$ and $v =
-\infty$, and is obviously extendible to $u > 0$, we find that the
effect of the perturbation is twofold (Fig. 4). First, the two
apparent null central singularities $u = 0$ and $v = -\infty$ are
replaced by two genuine spacelike central causal singularities
\cite{BTZ} $u + (M/4)v = 0$. Second, the two null apparent horizons,
displaced to \ba r_{,v} & = & -u^2(\rho' + (M/4)u^{-2}(x\rho)') =
\rho^2(\frac{u^2}{4l^2}-\frac{M}4) = 0, \nonumber \\ r_{,u} & = &
-(\rho + x\rho' + (M/4)u^{-2}(-x\rho+x(x\rho)')) =
-\rho^2(1-\frac{Mv^2}{16l^2}) = 0, \nonumber \\ \ea effectively shield
the final and initial singularities.

While the situation here is static, rather than dynamical, this
formation of apparent horizons in the linearized approximation (here
exact) is formally similar to that observed in critical collapse.
Recall that near-critical collapse is characterized by a critical
exponent $\gamma$, defined by the scaling relation \be\lb{scaling} Q
\propto |p - p^*|^{s\gamma}, \ee for a quantity $Q$ with dimension $s$
depending on a parameter $p$ (with $p = p^*$ for the critical
solution). In the present case, it has been shown \cite{cai} that the
BTZ vacuum is a critical point of the spinless BTZ black hole, and
that the phase transition from extremal ($M = 0$) to nonextremal BTZ
black holes is second order, so that a similar scaling relation
applies.   Choosing for $Q$ the radius \be\lb{scaling0} r_{AH} =
M^{1/2}l \ee of the apparent horizon ($s = 1$), and identifying $p -
p^*$ with the perturbation amplitude $M/4$, we obtain from
(\ref{scaling0}) the value of the BTZ critical exponent \be \gamma =
1/2, \ee in agreement with the value derived previously either from
the dynamical analysis of black-hole formation in 2+1 dimensions from
spherical dust collapse \cite{ps} or point particle collisions
\cite{bir}, or from the consideration of phase transitions from
non-extremal to extremal black holes \cite{cai} or from static to
rotating black holes \cite{kg}.

\setcounter{equation}{0}
\section{Scalar perturbations and black hole formation}

Let us return to the case $c^2 \neq 0$ where the scalar field does not
decouple. Eliminating now the functions $\ts$ and $\tp$ between the
three equations (\ref{pein01})-(\ref{pein04}), we arrive at the
fourth-order equation \be x^2\tr'''' -
\frac{2k-3c^2-7/2}{c^2+1}x\tr''' +
\frac{(k-c^2-1)(k-c^2-3/2)}{(c^2+1)^2}\tr'' \simeq 0, \ee which leads
to the secular equation \be\lb{sec}
p(p-1)\left(p-1-\frac{k}{c^2+1}\right)\left(p-\frac{k+c^2+1/2}{c^2+1}\right)
= 0.  \ee

Again, the root $p = 1 + k/(c^2+1)$ corresponds to the gauge mode
(\ref{gauge}). Discarding this mode, we obtain the general behavior of
the perturbations near $x = 0$ in terms of three integration constants
$A,\,B,\,C$: \ba & &\tr(x) \sim A +B(-x) + \L
C(-x)^{\frac{k+c^2+1/2}{c^2+1}},\lb{rti} \\ & &\ts(x) \sim - \frac{A}2
- \frac{c^2+2-k}{c^2+1}\L^{-1}B +O(x) \nonumber \\ & & \qquad \qquad
-\frac{(c^2+3/2)(k+c^2+1/2)}{(c^2+1)^2}C(-x)^{\frac{k-1/2}{c^2+1}},
\lb{sti}\\ & & \tp(x) \sim \frac{(2-k-c^2)}{2c}A +
\frac{c^2+2-k}{c(c^2+1)}\L^{-1}B + O(x) \nonumber \\ & & \qquad \qquad
+\frac{(c^2+3/2)(k+c^2+1/2)}{c(c^2+1)^2}C(-x)^{\frac{k-1/2}{c^2+1}},
\label{pti}
\ea for $k \neq 1/2$ (in the degenerate case $k = 1/2$, the power law
$(-x)^{\frac{k+c^2+1/2}{c^2+1}}$ is replaced by a logarithm).

From these behaviors, we find that the smooth matching conditions (B1)
at $v = 0$ are satisfied if either \be\lb{sol1} A = 0,\quad B = 0
\quad (Re(k) > 1/2), \ee or \be\lb{sol2} A = 0, \quad k = c^2 + 2.
\ee The stronger condition (B2) of analyticity in $v$ then implies, in
the first case, that $k = n(c^2 + 1) + 1/2$, where $n$ is a positive
integer, and, in the second case, that $C=0$. So the boundary
conditions (B1) and (B2) have two solutions: \ba & a) \quad A = B = 0,
\quad & k = n(c^2+1)+1/2, \lb{sola} \\ & b) \quad A = C = 0, \quad & k
= c^2 + 2. \lb{solb} \ea

When $u \to 0$ with $v$ held fixed, the corresponding metric
perturbations $(-u)^{(1-k)/(c^2+1)}\tr$ go to zero (in case a)) or to
a finite limit (in case b)), so that in all cases the perturbed $r$
has a finite limit and the boundary condition (A) is automatically
satisfied. If on the other hand one takes the limit $u \to 0$ with $x$
held fixed, then in all cases $r$ diverges as
$(-u)^{(1-k)/(c^2+1)}$. In case b) the divergence is the same (of the
order $(-u)^{-1}$) as in the BTZ case, while it is weaker in case a)
with $n = 1$, and stronger in case a) with $n \ge 2$. So a boundary
condition that $r$ does not diverge more than in the BTZ case when $u
\to 0$ with $x$ held fixed would select the eigenvalues $k = c^2 +
3/2$ (case a) with $n = 1$), or $k = c^2 + 2$ ( case b)).

Two of the three integration constants $A$, $B$, $C$ being now fixed,
the first-order perturbation is now completely fixed up to an
arbitrary scale, and (as discussed after (\ref{gauge})) within a gauge
transformation. So we now must check that the natural condition (C) at
the AdS boundary is indeed satisfied up to a gauge transformation. The
leading behaviour of the background at the AdS boundary ($x\to x_1$)
is, from Eqs. (\ref{bound0}),  \be \label{cdfg} \rho\simeq
\frac{\rho_1}{x-x_1}\>,\ \ \ e^{2\nu}\simeq
\left(\frac{4x_1}{\Lambda}\right)\frac{1}{(x-x_1)^2}\>,\ \ \ \psi
\simeq \psi_1\>.  \ee We again assume a power-law behavior
\be\lb{powq} \ts \sim b\bar{x}^q \ee ($\bar{x} = x-x_1$). Then
Eq. (\ref{pein2}), where $\tp$ can in first approximation be
neglected, gives \be q(q-1) = 2, \ee i.e. $q = -1$ or $q =
2$. Accordingly, Eq. (\ref{pein1}) reduces near $\bar{x} = 0$ to
\be\lb{asr} \tr'' - 2\bar{x}^{-2}\tr \simeq 4b\rho_1\bar{x}^{q-3}.
\ee If $q = -1$, the behavior of the solution is governed by the
right-hand side, i.e.  \be\lb{spu} \tr \sim b\rho_1\bar{x}^{-2}, \ee
which apparently violates the boundary condition (C) for $x \to x_1$.
However the behaviors (\ref{powq}) with $q = -1$ and (\ref{spu}) are
precisely those of the gauge mode (\ref{gauge}) near the AdS boundary,
so that they can be transformed away by a gauge transformation,
i.e. we can find a gauge\footnote{This gauge transformation is
equivalent to a shift in the position $x_1$ of the AdS boundary.} in
which the behavior (\ref{spu}) is replaced by \be\lb{asol} \tr \sim
\frac{E\rho_1}{x-x_1}, \ee ($E$ constant) which is consistent with the
boundary condition (C), and is an admissible small perturbation if its
amplitude is small enough, $¦E¦ \ll 1$. A finer analysis, where $\tp$
is not neglected in (\ref{pein2}), then leads to the behavior $\ts
\propto \bar{x}^2\ln|\bar{x}|$, and to the behavior (\ref{asol}) for
$\tr$.

Black hole formation is characterized by the appearance of a central
singularity hidden by an apparent horizon. The definition of these
notions can be quite ambiguous in linearized perturbation theory. For
our present purpose we will consider, rather than the centre $r = 0$,
the coordinate singularity of the perturbed metric  \be\lb{csing}
\sqrt{g} = e^{2\s}r = (-u)^{\a}e^{2\nu(x)}[\rho(x) +
(-u)^{-\ka}(\tr(x) + 2\rho(x)\ts(x))] = 0.   \ee  Let us discuss the
solution of this equation near $x = 0$, considering separately the two
cases a) and b). In case a) ($A=B=0$), we see from Eqs. (\ref{rti})
and (\ref{sti}) that $\tr+2\rho\ts$ is dominated by $2\ts$, so that
the coordinate singularity develops for \be\lb{csinga} (-u)^{\ka} =
-2\ts \simeq 2\frac{(c^2+3/2)(k+c^2+1/2)}{(c^2+1)^2}C
(-x)^{\frac{k-1/2}{c^2+1}}, \ee which may be rewritten as \be
(-u)^{\frac1{2(c^2+1)}} \propto Cv^n.  \ee For the lowest value $n =
1$ ($k = c^2+3/2$), we see that if $C > 0$ there is (to the order
considered) no coordinate singularity for $v < 0$, while a spacelike
coordinate singularity develops for $v > 0$. Note that the
consideration of the centre $r = 0$ would have given quite different
results, namely an eternal timelike centre at $(-u)^{\ka} = -\tr = -\L
Cx^2$ (for $n = 1$). The location of the trapping horizon is given by
\be\lb{aphor} r_{,v} = (-u)^{\frac{c^2+2}{c^2+1}}(\rho' +
(-u)^{-\ka}\tr') = 0, \ee or \be (-u)^{\ka} \simeq
2\frac{(c^2+2)(k+c^2+1/2)}{(c^2+1)^2}C (-x)^{\frac{k-1/2}{c^2+1}} \ee
near $x = 0$. For $n=1$, $C > 0$, this spacelike trapping horizon
develops for $v > 0$, just as the spacelike coordinate singularity
(\ref{csinga}) which it shields for all $c^2$ (Fig. 5). While we are
unable to show the formation of a curvature singularity without a
nonperturbative analysis, it is suggestive that this formation of a
shielded coordinate singularity is precisely what would  be expected
for a genuine curvature singularity. As the AdS boundary $x = x_1$ is
approached, it follows from the asymptotic behavior $\tr \sim E\rho +
F$ that both the coordinate singularity and the trapping horizon
approach the AdS boundary for a constant value of $u$, with a finite
limit for the radius of the apparent horizon \be\lb{limrad} r_{AH} =
\left(-\frac{\rho'}{\tr'}\right)^{1/k}\left(\rho-\rho'\frac{\tr}{\tr'}\right)
\quad \to \left(\frac FE\right)E^{1/k}.  \ee Note that it is essential
for this trapping horizon to form that the cosmological constant be
non-zero. For $\L = 0$, Eq. (\ref{csinga}) for the coordinate
singularity becomes exact, while on the other hand $\rho = 1$ and $\tr
= 0$ everywhere so that $r_{,v}$ is identically zero.

In case b) ($A = C = 0$, $k = c^2+2$), a finer analysis leads to the
behaviours (consistent with the exact solution (\ref{btzp})) near $x =
0$ \ba \tr  & \sim & -B(x + \frac{\L}4(1+c^2)x^2 + \cdots), \\ \ts &
\sim & -B(\frac{c^2}2 x + \cdots), \\ \tp & \sim & B(cx + \cdots).
\ea From these, it follows that that if $B < 0$, there is no
coordinate singularity for $v < 0$, a spacelike coordinate singularity
developing for $v > 0$ according to \be\lb{csingb} (-u)^{\a} \simeq
-B(1+c^2)v.  \ee  On the other hand, the location of the apparent
horizon is given near $x =0$ by  \be\lb{aphorb}
(-u)^{\frac{c^2+2}{c^2+1}} = - \frac{\tr'}{\rho'} \simeq \bar{B}
(1+(\gamma-\beta)x),  \ee  with $\bar{B} = 2B(c^2+2)/\L(c^2+1) > 0$,
$\beta = -\L(c^2+1)/2$, $\gamma = -\L(c^2+1)(4c^2+9)/2(2c^2+3)^2$, $0
< \gamma < \beta$ (the first order behaviour of $\rho'$ has been
computed from Eq. (\ref{einx1})). This  apparent horizon preexists the
singularity (Fig. 6), its radius varying as \be r_{AH} \simeq
[\bar{B}(1+(\gamma-\beta)x)]^{1/k}(1+O(x^2)), \ee i.e. increasing with
decreasing $x$, as suits a collapsing configuration. The behaviour of
the apparent horizon in the region $x>0$ is less clear. In particular,
it would seem to be asymptotic to the null line $u=0$ for some finite
value  $x=x_0$ (corresponding to $v\to-\infty$) , but at this point
$r_{AH}$ would diverge, in contradiction with our condition (A) for
the validity of the linear perturbation analysis. A more refined
calculation, including higher order perturbations, would be needed in
order to clarify this issue.

Now we inquire whether the apparent horizon of the perturbed spacetime
is regular. The non-constant part of the Ricci scalar (\ref{risca})
is, to first order, the product of three factors, \ba\lb{riscap} & R +
6\L = & 4e^{-2\s}\phi_{,u}\phi_{,v} =
\,4e^{-2\nu}\,[1-2(-u)^{-\ka}\ts]\cdot \nonumber \\ & &
[-\frac{c}{c^2+1} + x\psi' + (-u)^{-\ka}(\ka\tp+x\tp')]\cdot
\nonumber\\ & & [\psi'+(-u)^{-\ka}\tp']\,.  \ea Possible curvature
singularities for $v = 0$ having being excluded by the analyticity
condition (B2), perturbative curvature singularities can only arise
from the divergence of any factor of (\ref{riscap}) on the original
null singularity $u = 0$ \footnote{This again is due to a deficiency
of the perturbative approach: the perturbative coordinate singularity
due essentially to the vanishing of $e^{2\s}$ may, as mentioned above,
be a more correct indicator of a curvature singularity.}. Leaving
aside case b), where as mentioned above it is not clear whether the
apparent horizon intersects the line $u = 0$, we concentrate on case
a). The perturbative Ricci scalar (\ref{riscap}) evaluated on the
apparent horizon (\ref{aphor}) is \ba\lb{riscaph} & R_{AH} + 6\L = &
4e^{-2\nu}\,[1+2\rho'\frac{\ts}{\tr'}]\cdot \nonumber \\ & &
[-\frac{c}{c^2+1} + x\psi' -
\rho'\frac{\ka\tp+x\tp'}{\tr'}]\,[\psi'-\rho'\frac{\tp'}{\tr'}]\,.
\ea The last factor $\phi_{,v}$ diverges as $x^{-1}$ when $x \to 0$,
while the other factors go to finite limits, so that the apparent
horizon will generically be singular at its onset $x = 0$, unless the
product $e^{-2\s}\phi_{,u}$ accidentally vanishes, \be
e^{-2\s}\phi_{,u}|_{AH}(x=0) = 0.  \ee

We suggest that this condition of ``regularity at birth'' of the
apparent horizon is the crucial boundary condition (D) which will
determine the eigenvalue of $k$. Specifically, for $x \to 0$,
\be\lb{riscaph0} R_{AH} + 6\L \propto \left(1 +
\frac{c^2+3/2}{c^2+2}\right)\left(c -
\frac{(2k-1/2)(c^2+3/2)}{2c(c^2+2)}\right)x^{-1}.  \ee The first
factor is positive, while the second factor vanishes for the value
\be\lb{k0} k_0 = \frac{1}{4} + c^2\frac{c^2+2}{c^2+3/2}.  \ee
Unfortunately, it is easy to see that $k_0 < c^2 + 3/2$ for any $c$,
so that the analyticity condition (B2) is not satisfied and
consequently the Ricci scalar diverges on the line $v =
0$. Nevertheless we believe that the mechanism proposed here for the
determination of $k$ is basically correct, and that it is its
implementation in the linear approximation which is responsible for
the problem just mentioned. Indeed, a way to overcome this problem is
to play with the ambiguities of the linear approximation and linearize
the product $e^{-2\s}\phi_{,u}$ {\em before} evaluating it on the
apparent horizon. Then, (\ref{riscaph0}) is replaced by
\be\lb{riscaph1} R_{AH} + 6\L \propto \left(c +
\frac{c(c^2+3/2)}{c^2+2} -
\frac{(2k-1/2)(c^2+3/2)}{2c(c^2+2)}\right)x^{-1}, \ee which vanishes
for the value \be\lb{k1} k_1 = \frac{2c^4+(15/4)c^2+3/8}{c^2+3/2}.
\ee This is smaller than $(2c^2+5/2)$ (the value (\ref{sola}) for
$n=2$), while the equation \be k_1 = c^2+3/2 \ee ($n=1$) has only one
real solution $c^2 \simeq 1.045$, leading to $k \simeq 2.545$. While
the linearization trick we have used to obtain this value is hard to
justify, it is nevertheless suggestive that we have found a set of
boundary conditions (B1), (B2), and (D) which determine a single value
for the couple ($c,\,k$), i.e. a unique critical solution with a
unique growing mode, and that this value $c^2 \simeq 1.045$ is very
close to the value $c^2 = 1$ suggested as a critical value by the
global analysis of our family of CSS solutions (Figs. 1 and 2). Let us
also recall that Garfinkle \cite{gar} has found that the solution
(\ref{gar1}) is in good agreement with the numerical results of
\cite{CP} at an intermediate time for the special value $c^2 = 7/8 =
0.875$, which is also close to 1.

From the dependence (\ref{limrad}) of the limiting horizon radius on
the perturbation amplitude (proportional to the sole non-vanishing
integration constant $C$ or $B$), the critical exponent is $\gamma = 1/k$. For
the preferred value $c^2 \approx 1$, this leads to $\gamma \approx
2/5$ \cite{crit} in case a) with $n = 1$, or $\gamma \approx 1/3$ in
case b).

\setcounter{equation}{0}
\section{Discussion}

The reason for writing this paper has been twofold. In the first part
we have extended previous works \cite{gar,crit} motivated by the quest
for the critical solution underlying the  collapse of a massless and
minimally coupled scalar field in 2+1 dimensional anti-de Sitter
spacetime. In particular, we have constructed a family of exact
solutions with CSS behavior near the singularity and AdS behavior at
spatial infinity. This one-parameter ($c$) family of time-dependent
solutions interpolates between the static vacuum solution ($c^2=0$)
and the massive BTZ black hole ($c^2=\infty$), with the critical value
$c^2=1$ lying at the boundary between two different singularity types
(point singularity, and line singularity).

These solutions served as the starting point for the subsequent linear
perturbation analysis.  A crucial question is that of the choice of
the correct boundary conditions to be imposed along the null
boundaries $u=0$, $v=0$, and the AdS boundary $x=x_1$.  These boundary
conditions have been ``tested'' in the simple case $c^2=0$, where the
scalar field decouples and the linear perturbations describe exactly
the small mass static BTZ black hole. In this respect, the essential
difference with our previous work \cite{crit} is that we now recognize
that the condition (C) that the perturbations do not blow up at the
AdS boundary is always satisfied up to gauge transformations; we can
then impose the natural smooth matching conditions (B1)
(Eq. (\ref{bound2})) on the boundary $v=0$, instead of the weaker
condition $\tr(0) = 0$ used in \cite{crit} (resulting in significantly
different behaviors of the coordinate singularity and apparent
horizon). We also considered a new and crucial condition (D), stating
the ``regularity at birth'' of the apparent horizon. Fulfilment of the
selected boundary conditions then gives rise to two distinct growing
modes a) (Eq. (\ref{sola}) with $n=1$) and b) (Eq. (\ref{solb})). The
application of condition (D) is very suggestive because, together with
the other boundary conditions it selects, for the mode a), a unique
value for both the eigenvalue $k$ and the parameter $c$.

The ensuing physical picture, along with the ambiguities inherent to
the linear perturbations approximation, is also discussed. For the
mode a) the singularity and the apparent horizon appear simultaneously
on the null line $u = 0$ at the time $v=0$ and evolve in the region
$v>0$ in a physically meaningful way. On the other hand, for the mode
b) (which is the extension to $c^2 \neq 0$ of the single BTZ mode
found for $c^2=0$), the singularity still appears for $v = 0$, but the
apparent horizon seems to be eternal, as in the case of the static BTZ
black hole. This formal solution cannot describe actual gravitational
collapse with regular initial conditions. Indeed, we conjecture
\cite{critmosc} that the solution b) is actually a linearization of
the singular static solution (\ref{cicca}). We would like to argue
that ultimately the boundary conditions (apart from natural
requirements on the consistency of the linear approximation) should be
chosen so as to optimize the agreement of the perturbative picture
with realistic gravitational collapse. On such a basis the mode b)
should be excluded as unphysical. This would mean that there is
exactly one growing mode a), showing that the background quasi-CSS
solution is indeed the critical solution for gravitational collapse.

An objection against this view is that the quasi-CSS solutions presented in
this paper do not have a regular timelike origin, while all
near-critical collapse scenarios do evolve about a regular timelike
origin. However we argue that a critical solution, which lies in the
boundary surface separating different collapsing regimes, does not
necessarily share with the generic collapsing configurations a regular
central behavior, i.e. the critical solution can lie at the boundary 
of the set of generic
collapsing solutions without belonging to the set. Indeed, further
investigations currently under
progress seem to show that certain perturbations of our quasi-CSS
solutions do present a regular timelike center, and that, moreover, 
a subset of these perturbations correspond
precisely to extended (to $\Lambda < 0$) Garfinkle solutions. This would
suggest that our quasi-CSS solutions, with singular center, could be
exact critical solutions, with near-critical solutions well described
by the extended Garfinkle solutions.

We finally turn to the calculation of the critical exponent
$\gamma$. Numerically, Pretorius and Choptuik \cite{CP} found $\gamma$
in the range  $1.15 < \gamma < 1.25$, while Husain and Olivier
\cite{HO} estimated $\gamma \sim 0.81$.  Choosing the critical value
$c^2 = 1$ (which is also motivated by the application of boundary
condition (D)), we obtain for the mode a) the value $\gamma = 0.4$,
which differs significantly from both numerical values. We cannot at
present give a reason for this disagreement (which recalls the disagreement
between the numerical \cite{chop} and analytical \cite{fro}
determinations of the critical exponent for scalar field collapse in
3+1 dimensions).  Despite this shortcoming, we find it rather satisfactory to
have been able to reproduce from this solution a number of
characteristic features of critical collapse and black hole formation.

\bigskip
\noindent {\bf\large Acknowledgments}

\medskip
We would like to thank Carsten Gundlach and Sean A. Hayward for discussions.
We also thank the referee for his constructive comments.

\newpage

\begin{figure}
\centerline{\epsfxsize=300pt\epsfbox{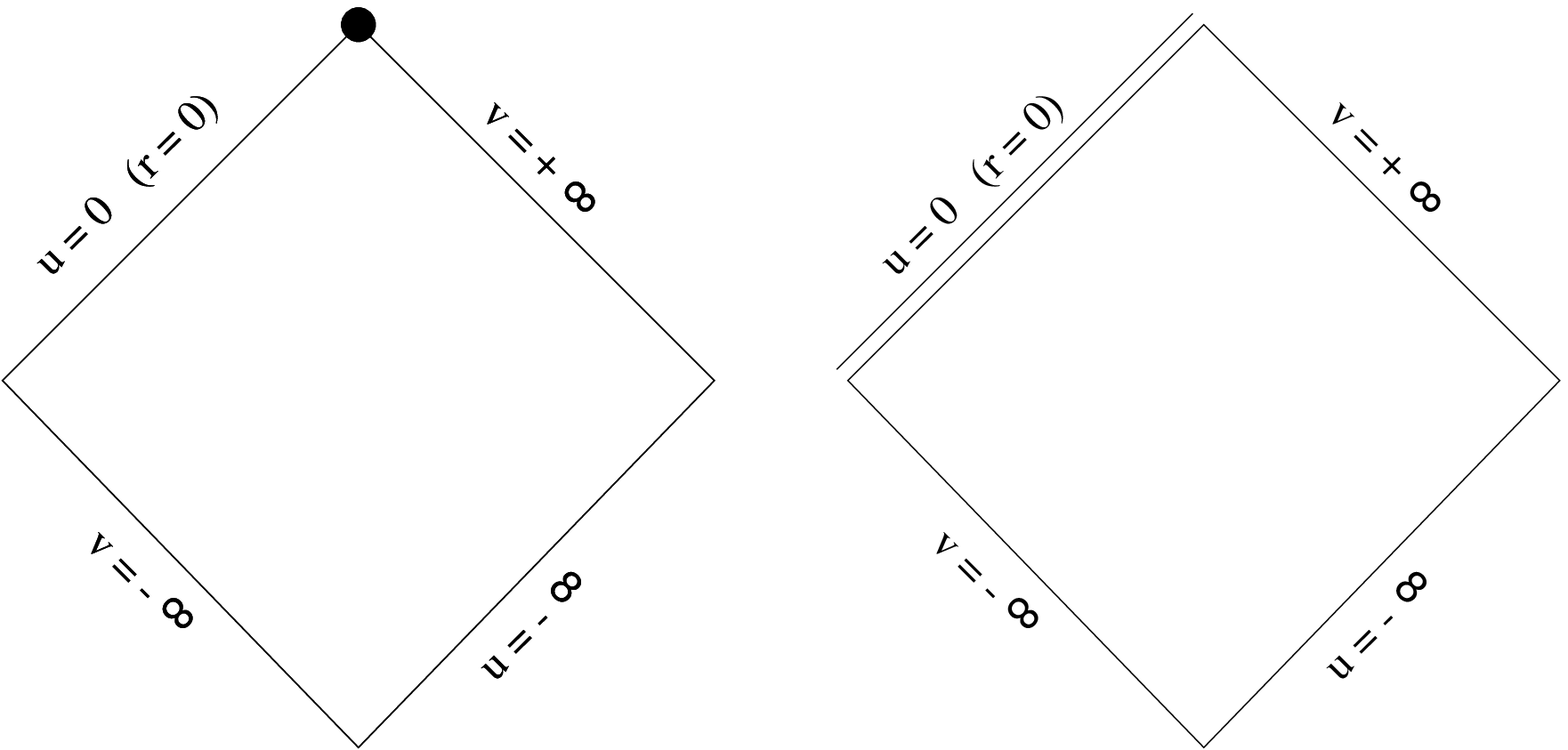}} \caption{Penrose
diagram of the CSS solutions eq. (\ref{new2}) ($\L= 0$) a) for $c^2 \le
1$, with a point singularity (dot) at $u=0$, $v=+\infty$; b) for $c^2
> 1$, with a line singularity (double line) at $u = 0$.}
\end{figure}

\begin{figure}
\centerline{\epsfxsize=200pt\epsfbox{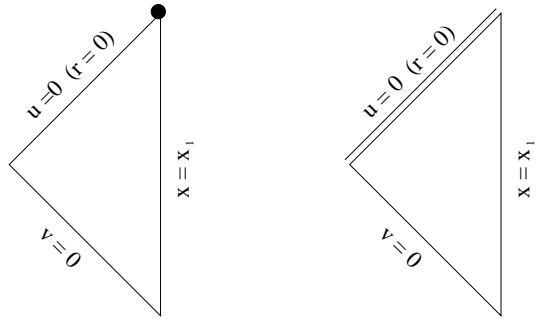}} \caption{Penrose
diagram of the extended quasi-CSS solutions ($\L < 0$) a) for $c^2 \le
1$; b) for $c^2 > 1$.}
\end{figure}

\vfill
\newpage

\begin{figure}
\centerline{\epsfxsize=200pt\epsfbox{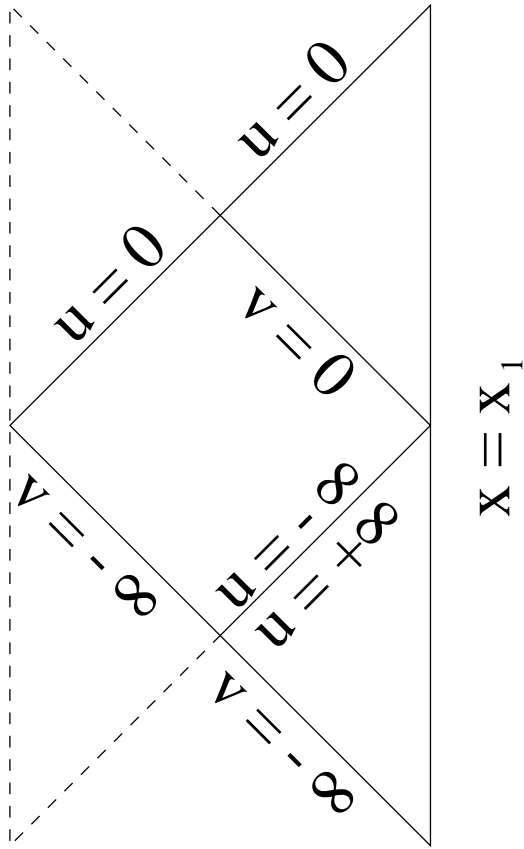}}
\caption{Penrose diagram of the vacuum solution (\ref{vac2}). The
analytic continuation from $u = -\infty$ to $u = +\infty$ is achieved
with the map (\ref{new0i}).}
\end{figure}

\begin{figure}
\centerline{\epsfxsize=200pt\epsfbox{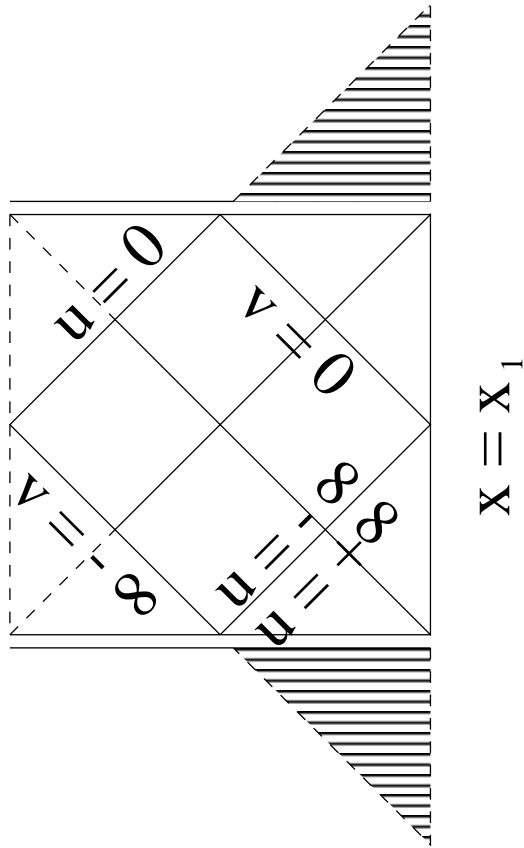}}
\caption{Penrose diagram of the BTZ black hole (\ref{btz5}). The two
spacelike central singularities (double lines) are shielded by two
null horizons (the diagonals).}
\end{figure}

\begin{figure}
\centerline{\epsfxsize=200pt\epsfbox{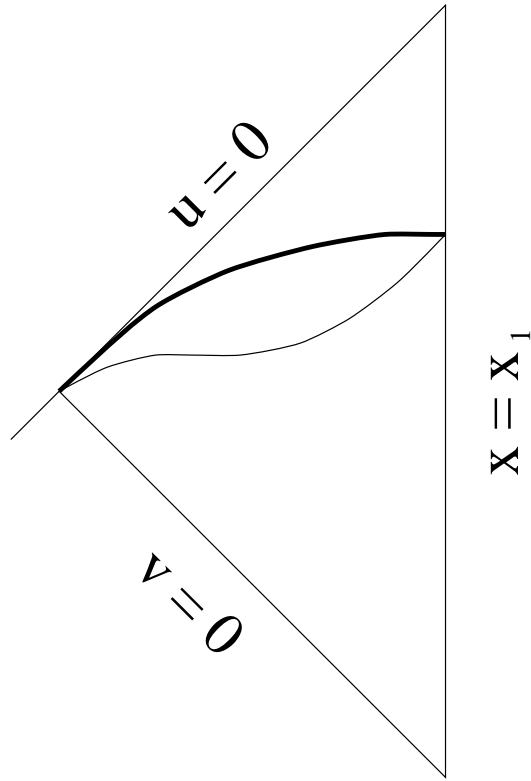}}
\caption{Scalar perturbations with $k = c^2 + 3/2$ (case a)). The
spacelike coordinate singularity (thick curve) and apparent horizon appear
simultaneously at v = 0.}
\end{figure}

\begin{figure}
\centerline{\epsfxsize=200pt\epsfbox{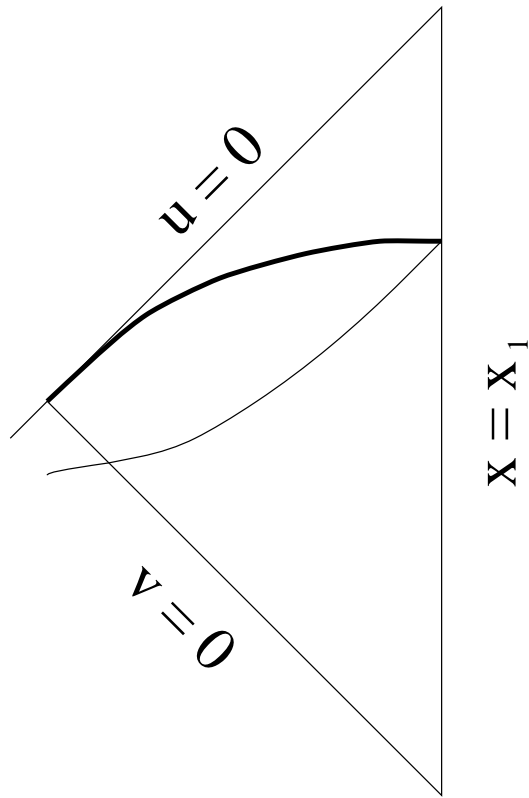}}
\caption{Scalar perturbations with $k = c^2 + 2$ (case b)). The
apparent horizon prexists the spacelike singularity.}
\end{figure}

\end{document}